# Probing the minigap in topological insulator-based Josephson junctions under radio frequency irradiation[*]

Guang Yang(杨光)[1,2], Zhaozheng Lyu(吕昭征)[1,2], Xiang Zhang(张祥)[1,2], Fanming Qu(屈凡明)[1,3] and Li Lu(吕力)[1,2,3,4]

[1]*Beijing National Laboratory for Condensed Matter Physics, Institute of Physics, Chinese Academy of Sciences, Beijing 100190, China*
[2]*School of Physical Sciences, University of Chinese Academy of Sciences, Beijing 100049, China*
[3]*CAS Center for Excellence in Topological Quantum Computation, University of Chinese Academy of Sciences, Beijing 100190, China*
[4]*Beijing Academy of Quantum Information Sciences, Beijing 100193, China*

Recently, a contact-resistance-measurement method was developed to detect the minigap, hence the Andreev bound states (ABSs), in Josephson junctions constructed on the surface of three-dimensional topological insulators (3D TIs). In this work, we further generalize that method to the circumstance with radio frequency (rf) irradiation. We find that with the increase of rf power, the measured minigap becomes broadened and extends to higher energies, in a way similar to the rf power dependence of the outer border of the Shapiro step region. We show that the corresponding data of contact resistance under rf irradiation can be well interpreted by using the resistively shunted Josephson junction model (RSJ model) and the Blonder-Tinkham-Klapwijk (BTK) theory. Our findings could be useful when using the contact-resistance-measurement method to study the Majorana-related physics in topological insulator-based Josephson junctions under rf irradiation.

[*] Project supported by the National Basic Research Program of China from the MOST grants 2016YFA0300601, 2017YFA0304700, and 2015CB921402, by the NSF China grants 11527806, 91221203, 11174357, 91421303, 11774405, and by the Strategic Priority Research Program B of the CAS grant No. XDB07010100, XDB28000000.

## 1. Introduction

In 2008, Fu and Kane proposed that the superconducting proximity effect between an s-wave superconductor (S) and the surface of a 3D TI can induce p-wave-like superconductivity and host Majorana bound states (MBSs) in vortex cores or in proximity-type S-TI-S Josephson trijunctions.[1] In more detail, the electron-like and hole-like ABSs in single S-TI-S Josephson junctions are predicted to have $4\pi$-period energy phase relations (EPRs), and the 1D Majorana modes are predicted to be fully decoupled when the phase difference reaches $\pi$, resulting in the complete close of minigap between electron-like and hole-like ABSs. In Josephson trijunctions on 3D TIs, furthermore, MBS is predicted to exist at the center of the trijunction over extended regions in phase space. In past years, experimental efforts have been paid to search for $4\pi$-period current phase relations (CPRs) which are direct consequences of $4\pi$-period EPRs. Some signatures, such as skewed CPR,[2-4] missing of odd Shapiro steps,[5-8] etc., have been discovered. Meanwhile, a contact-resistance-measurement method has also been developed for directly probing the EPRs in the junction area.[9-11] A linear closing behavior of minigap as signature of $4\pi$-period EPRs in single S-TI-S junctions, and the complete close of minigap at the center of S-TI-S trijunctions as evidence of MBS, have been observed.[10, 11]

However, MBSs suffer poisoning from quasiparticle fluctuations. To further study the MBSs, such as to verify their non-Abelian statistics and to perform quantum computation in the future, one would require fast braiding/fusion operations and readout of MBSs based on rf techniques. One of the ways of going to rf frequencies is to shine the S-TI-S junctions with rf irradiation.

For the purpose of studying the minigap of the surface state in S-TI-S junctions in the presence of rf irradiation, in this work we generalized the contact-resistance-measurement method to the circumstance with rf irradiation. We find that this method still works --- the measured data of contact resistance under rf irradiation can be well understood and numerically simulated by using the RSJ model[12] and the BTK theory.[13]

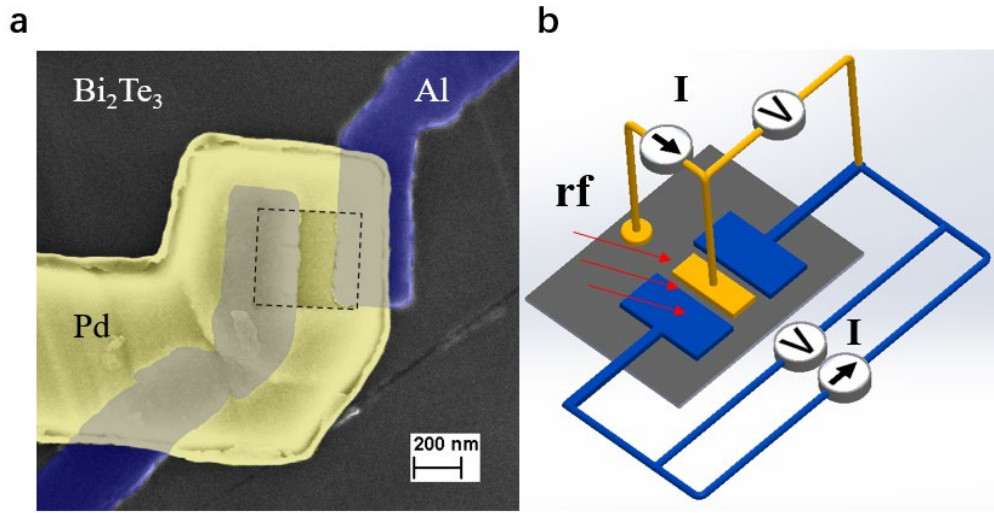

Fig. 1. (color online) Device structure and measurement configuration. (a), False-color scanning electron microscopy image of the proximity-type Josephson junction. A Josephson junction with Al electrodes (blue) was fabricated on the surface of a $Bi_2Te_3$ flake (gray). The normal-metal Pd electrode (yellow) was deposited to contact the junction area through a window (marked by the dashed rectangle) on insulating mask made of overexposed PMMA. Note that the surface of Al electrodes was oxidized in situ before evaporating Pd. (b), Schematics of the device under rf irradiation. The configurations for contact resistance measurement (yellow) and for Josephson supercurrent measurement (blue) are illustrated.

## 2. Experiment

Figure 1a shows the false-color scanning electron microscopy (SEM) image of the device. The two superconducting Al electrodes couple with each other through a $Bi_2Te_3$ flake (~100 nm thick), to form a Josephson junction of length $L$~153 nm and width $W$~484 nm. The surfaces of Al electrodes are oxidized in situ after deposition. An additional insulating layer of overexposed polymethyl methacrylate (PMMA) was fabricated to cover the device except for a window at the junction area (marked by the dashed rectangle in Fig. 1a). Through this window, a normal-metal Pd electrode is further deposited to electrically contact with the TI surface in the junction area for contact resistance measurement. Figure 1b shows the schematics of the device, together with the configurations for Josephson supercurrent measurement (blue circuits) and contact resistance measurement (yellow circuits). The measurements were carried out in a dilution refrigerator with a base temperature of ~10 mK. The rf irradiation was applied via a coaxial cable with an open end as an antenna near the device.

In the following, we will first present and discuss the data of $I_J - V_J$ curves (where $I_J$ is the current passing from one superconducting electrode to another, and $V_J$ is the voltage drop across the Josephson junction), and then present and discuss the data of contact resistance between Pd electrode and Bi$_2$Te$_3$ under rf irradiation.

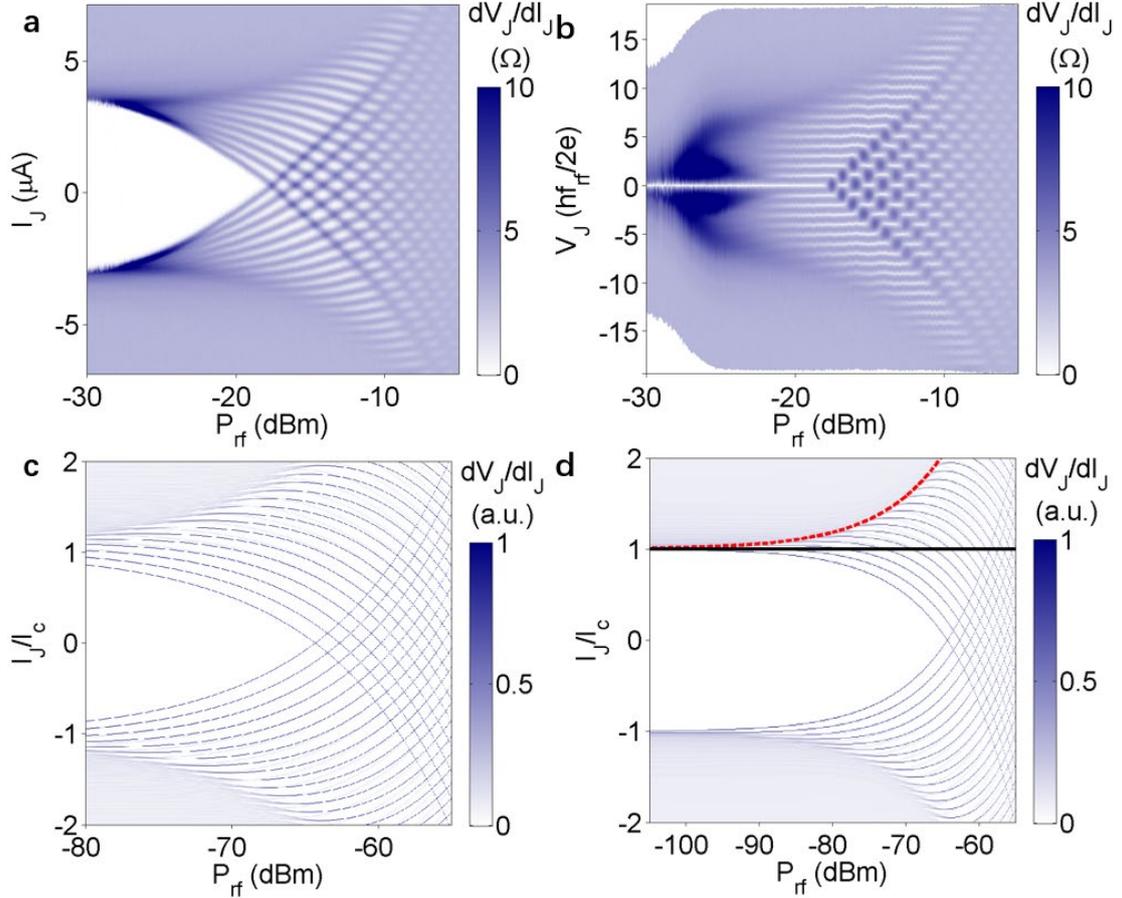

Fig. 2. (color online) Shapiro maps in an Al-Bi$_2$Te$_3$-Al junction. (a), Differential resistance $dV_J/dI_J$ as a function of rf power $P_{rf}$ and direct bias current $I_J$, measured at rf frequency $f_{rf} = 0.5$ GHz. (b), The same data as in (a), but replotted as a function of rf power $P_{rf}$ and direct voltage $V_J$ across the junction in the unit of $hf_{rf}/2e$. Evenly spaced Shapiro steps can be seen. (c), Simulated $dV_J/dI_J$ based on the RSJ model. (d), The same simulated $dV_J/dI_J$ as shown in (c), but in a wider range of rf power. The black line represents the critical supercurrent $I_c$, and the red curve represents the sum of the critical supercurrent $I_c$ and the amplitude of rf-driving current $I_{rf0}$.

## 3. Results and discussion

In the presence of rf irradiation, it is known that there will be Shapiro steps on the $I_J - V_J$ curves of a Josephson junction.[12] Figure 2a shows the measured differential resistance $dV_J/dI_J$ as a function of $I_J$ and rf power $P_{rf}$ applied to the antenna. Figure 2b shows the same data as in Fig. 2a, but converted and replotted as a function of $V_J$ and $P_{rf}$. The horizontal lines correspond to well defined Shapiro steps in $I_J - V_J$ curves at voltages $V_{Jn} = nhf_{rf}/2e = n \times 1.04$ μV, where $h$ is the Planck constant, $f_{rf} = 0.5$ GHz is the rf frequency, $e$ is the electron charge, and $n = 0, \pm 1, \pm 2....$

Usually, the Josephson supercurrent has a $2\pi$-period CPR, which leads to the appearance of ordinary Shapiro steps on $I_J - V_J$ curves in the presence of rf irradiation. If the supercurrent contains a $4\pi$-period component, however, the odd Shapiro steps might disappear, leading to the so called fractional Josephson effect.[14] In some S-TI-S junctions, the missing of the odd Shapiro steps has been observed.[5-7] In our devices, however, Fig. 2b shows that all the Shapiro steps are present. It indicates that the dominant supercurrent in our devices has a $2\pi$-period CPR, flowing presumably through the bulk of Bi$_2$Te$_3$ flake,[10] and that the $4\pi$-period component of supercurrent in our device is negligibly small.

The $I_J - V_J$ curves with rf irradiation can be simulated by solving the following equation based on the RSJ model[12]:

$$I_J + I_{rf0} \sin(2\pi f_{rf} t) = I_c \sin(\phi(t)) + \frac{d\phi(t)}{dt} \times \frac{1}{4\pi e R}, \tag{1}$$

where $I_J$ is the direct current, $I_{rf0} \propto \sqrt{P_{rf}}$ is the amplitude of alternating current caused by rf irradiation, $I_c = 3.46$ μA is the critical supercurrent, $\phi(t)$ is the superconducting phase difference, $R = 2.8$ Ω is the resistance of the junction in the normal state. The details of numerical simulation can be found in the supplementary materials. Figure 2c shows the simulated $P_{rf}$ and $I_J$ dependence of $dV_J/dI_J$. The result reproduces the main features of the experimental data in Fig. 2a.

With increasing rf power, the region containing Shapiro steps in Fig. 2 extends to higher currents/voltages. The border of this region roughly coincides with the red dashed curve in Fig. 2d which represents the sum of the critical supercurrent $I_c$ (the black line) and the amplitude of the rf-driving current $I_{rf0}$. It reflects that the Josephson junction can still have the chance to mode-lock with the rf driving frequency (i.e., forming the Shapiro steps), as long as the total bias current $I_J + I_{rf0}\sin(2\pi f_{rf}t)$ has the chance to be smaller than the critical supercurrent $I_c$ within each oscillation period.[14]

In the next, let us present and discuss the data of the contact resistance $dV_b/dI_b$ across the Pd-Bi$_2$Te$_3$ interface. Figure 3a shows the $dV_b/dI_b$ as a function of rf power $P_{rf}$ and bias current $I_b$. The vertical line cuts of Fig. 3a at low rf powers take basically the same line shape as the red curve shown in Fig. 3b, with a superconductivity-related dip-peak-dip structure due to the existence of a minigap on the TI surface of the S-TI-S junction.[10] Such a line shape can be well understood within the framework of the BTK theory.[10, 13] The black curve in Fig. 3b is the BTK fitting to the experimental data, with fitting parameters as follows: the minigap $\Delta_0 = 20.4\ \mu eV$, the effective number of channels of the contact $N = 25.5$, the barrier strength $Z = 0.66$ and the effective electron temperature $T = 0.1\ K$.

With increasing rf power $P_{rf}$, the dip-peak-dip structure on the vertical line cuts of Fig. 3a, shown as the red curves in Fig. 3d, gradually rounds up and spreads to higher $I_b$. The characteristic width of this structure, as indicated by the red dashed line in Fig. 3c, follows a similar trace as the outer border of the Shapiro step regions in Fig. 2a.

Theoretically, shining the device with rf irradiation will cause two effects on contact resistance measurement. First, it will generate a rf current $I_{rf0}\sin(2\pi f_{rf}t)$ flowing mainly through the bulk of the junction, by which influencing the time-dependent phase difference $\phi(t)$ of the Josephson junction, thus influencing the minigap $\Delta(t)$. Second, the rf irradiation will also generate a rf current passing through the Pd-Bi$_2$Te$_3$ interface, by which modifying the contact resistance measurement. Combining these two effects, the contact resistance can be simulated as follows, still by using the RSJ model and the BTK theory.

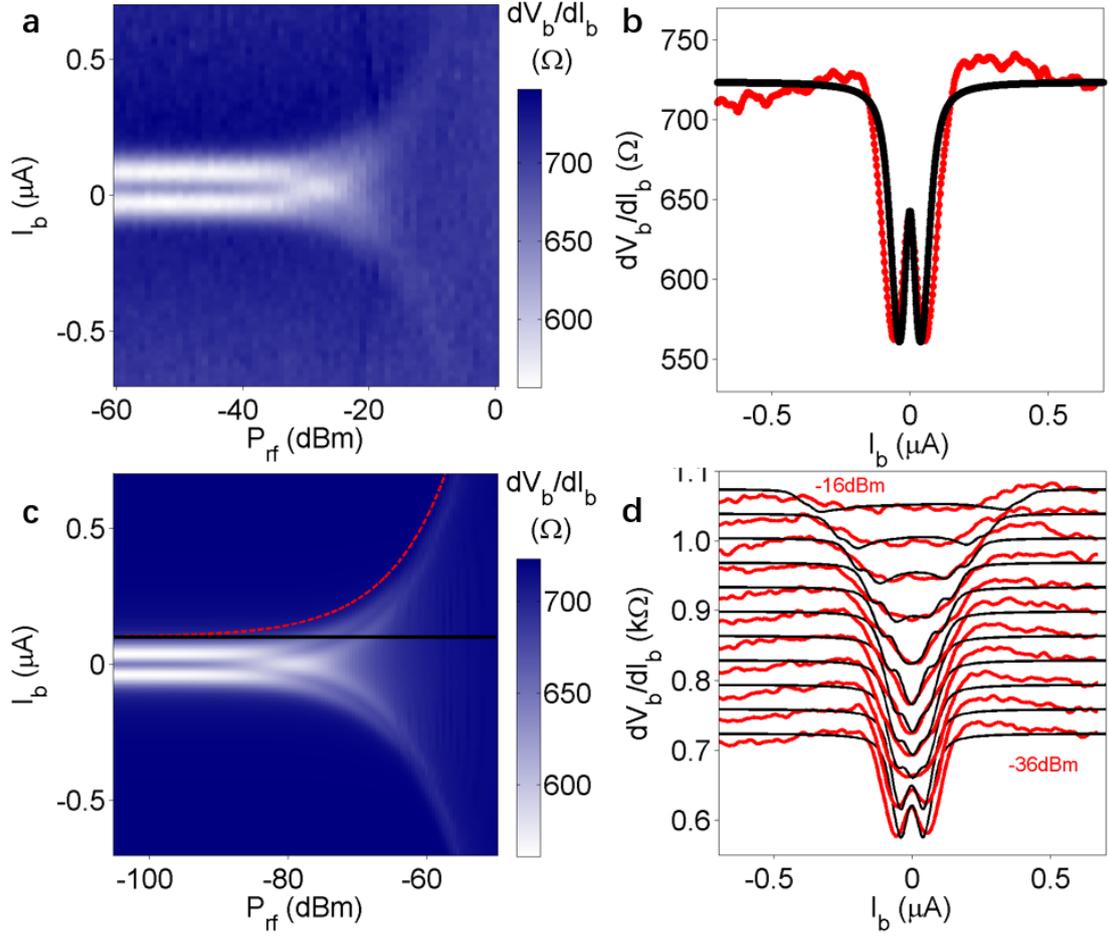

Fig. 3. (color online) Differential contact resistance in the junction. (a), The differential contact resistance $dV_b/dI_b$ across the Pd-$Bi_2Te_3$ interface as a function of rf power $P_{rf}$ and bias current $I_b$. (b), The $dV_b/dI_b$ as a function of bias current $I_b$ in the absence of rf irradiation (red curve), together with the BTK fitting (black curve). (c), The $dV_b/dI_b$ calculated from the RSJ model and the BTK theory. The black line represents a characteristic current $I_e$ at zero rf power, beyond which the signature of superconductivity disappears. The red dashed line represents the sum of $I_e$ and the amplitude of the rf driving alternating current $I'_{rf0} \propto \sqrt{P_{rf}}$ across the Pd-$Bi_2Te_3$ interface. (d), rf power dependence of the measured $dV_b/dI_b - I_b$ curves (red) and the simulated ones (black) (curves are shifted vertically for clarity).

In the presence of rf irradiation, the time-dependent minigap of the surface state should follow the $4\pi$-period form[10] and can be expressed as: $\Delta(t) = \Delta_0 |\cos(\phi(t)/2)|$, where the time-dependent phase difference $\phi(t)$ can be obtained by solving Eq. (1). On the other hand, the time-dependent total bias current passing through the Pd-Bi$_2$Te$_3$ interface is $I_\text{b} + I'_\text{rf0}\sin(2\pi f_\text{rf} t)$, where $I_\text{b}$ is the direct bias current and $I'_\text{rf0} \propto \sqrt{P_\text{rf}}$ is the amplitude of the rf driving current. We can obtain the instant value of $\text{d}V_\text{b}(\text{t})/\text{d}I_\text{b}(\text{t})$ by using the fitting parameters $\Delta_0$, $N$, $Z$ and $T$ obtained before. Then, the time average of $\text{d}V_\text{b}(\text{t})/\text{d}I_\text{b}(\text{t})$, $\text{d}V_\text{b}/\text{d}I_\text{b}$, can be numerically obtained. The details of the simulation can be found in the supplementary materials. The simulated results are plotted in Fig. 3c and also as the black curves in Fig. 3d at several different rf powers. The red dashed line in Fig. 3c represents the border of $I_\text{b} \leq I_\text{e} + I'_\text{rf0}$, where $I_\text{e}$ is a characteristic current beyond which there is no signature of superconductivity without rf irradiation. It can be seen that the rf driving current makes the measured minigap structure extend to higher bias current under rf irradiation. Nevertheless, the numerical simulations can still reproduce the measured data.

## 4. Conclusion

To conclude, we have examined and confirmed the validity of contact-resistance-measurement method for detecting the minigap in S-TI-S Josephson junction under rf irradiation. Although both the phase difference across the Al-Bi$_2$Te$_3$-Al junction and the minigap measurement process across the Pd-Bi$_2$Te$_3$ interface are all influenced by rf irradiation, the measured contact resistance can still be interpreted by using the RSJ model and the BTK theory. These results might be useful when using the contact-resistance-measurement method to study the minigap of S-TI-S junctions under rf irradiation and Majorana-related physics.

# Supplemental Material for "Probing the minigap in topological insulator-based Josephson junctions under radio frequency irradiation"

Guang Yang(杨光)[1,2], Zhaozheng Lyu(吕昭征)[1,2], Xiang Zhang(张祥)[1,2], Fanming Qu(屈凡明)[1,3] and Li Lu(吕力)[1,2,3,4]

[1]Beijing National Laboratory for Condensed Matter Physics, Institute of Physics, Chinese Academy of Sciences, Beijing 100190, China

[2]School of Physical Sciences, University of Chinese Academy of Sciences, Beijing 100049, China

[3]CAS Center for Excellence in Topological Quantum Computation, University of Chinese Academy of Sciences, Beijing 100190, China

[4]Beijing Academy of Quantum Information Sciences, Beijing 100193, China

**Contents**

1. The resistively shunted junction model
2. Simulating the Shapiro map
3. Simulation of the differential contact resistance

## 1. The resistively shunted junction (RSJ) model

A Josephson junction can be treated as an equivalent circuit with a pure Josephson junction J in parallel with a shunted resistor R, as shown below.[1]

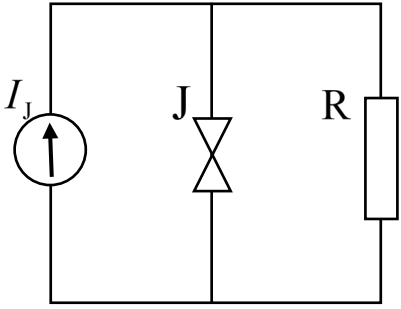

The current continuity equation for the circuit is:

$$I_\text{J} = \frac{V_\text{J}(t)}{R} + I_\text{c}\sin(\phi(t)),$$

where $I_\text{J}$ is the dc bias current, $R$ is the resistance of the Josephson junction in the normal state, $V_\text{J}(t)$ is the voltage across the resistor/Josephson junction and $I_\text{c}\sin(\phi(t))$ is the supercurrent in the Josephson junction (where $I_\text{c}$ is the critical supercurrent and $\phi(t)$ is the superconducting phase difference of the Josephosn junction).

In the Josephson junction, $d\phi(t)/dt = 2eV_\text{J}(t)/\hbar$, where $\hbar$ is the reduced Planck constant, so that the current continuity equation becomes:

$$I_\text{J}(t) = \frac{\hbar}{2eR} \times \frac{d\phi(t)}{dt} + I_\text{c}\sin(\phi(t)).$$

## 2. Simulating the Shapiro map

When the Josephson junction is exposed to radio frequency (rf) irradiation, an ac current $I_{rf}$ will be applied to the Josepshon junciton. The equivalent circuit now becomes:

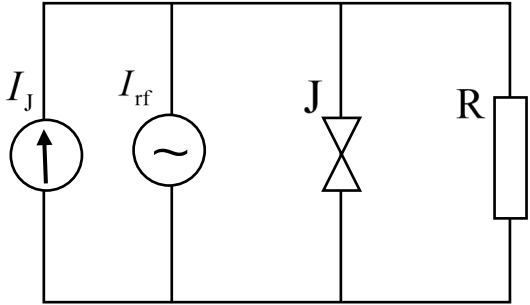

The current continuity equation now becomes:

$$I_J + I_{rf0}\cos(2\pi f_{rf}t) = \frac{\hbar}{2eR}\times\frac{d\phi(t)}{dt} + I_c\sin(\phi(t)),$$

where $I_{rf0}$ is the amplitude of the rf-driving ac current which depends on rf power $P_{rf}$, and $f_{rf}$ is the frequency of the rf irradiation. $R$ and $I_c$ can be measured by applying $I_J$ to reach the normal state. Since the $2\pi$-period supercurrent dominates in the Josephson junction, we always neglect the $4\pi$-period component in the current continuity equation.

In simulating the Shapiro map ($dV_J/dI_J$ vs. $I_J$ and $P_{rf}$, see Fig. 2c and 2d in the main manuscript), firstly the amplitude of the rf-driving ac current $I_{rf0}$ is related to the rf power $P_{rf}$ in the formula: $P_{rf}(dBm) = 10\log_{10}(I_{rf0}^2 R_0/1000)$, where $R_0$ is 50 Ω. At given $I_J$ and $I_{rf0}$, we can calculate the instant phase difference $\phi(t)$ by numerically solving the current continuity equation above. Then, the instant voltage $V_J(t) = \hbar/2e \times d\phi(t)/dt$ can be calculated, and the dc voltage $V_J$ can be further obtained by averaging $V_J(t)$ over time. In such a way, the $I_J-V_J$ curves as well as the $dV_J/dI_J$ vs. $I_J$ curves can be calculated. Lastly, the $dV_J/dI_J$ as a function of $I_J$ and $P_{rf}$, namely the Shapiro map, can be obtained.

## 3. Simulation of the differential contact resistance

We have proven that the Blonder-Tinkham-Klapeijk (BTK) theory can be used to describe the electron trsnsport processes across the interface between a normal metal and the surface of a three dimentional topological insulators (3D TIs) with proximity induced superconductivity.[2, 3] By fitting the red curve in Fig. 3b of the main manuscript, the differential contact resistance versus the bias current, with BTK theory, we can get the minigap in the absence of rf irradiation $\Delta_0 = 20.4\ \mu eV$, the effective number of conduction channels of the contact $N = 25.5$, and the barrier strength of the interface $Z = 0.66$. With these parameters, we can calculate the differential contact resistance $dV_b/dI_b$ as a function of mingap and bias current by using BTK theory. The results are shown below. We call it the the differential contact resistance map, or the R map. From this map we can extract the differetial contact resistance at any instant time if we know the instant values of the minigap and the bias current.

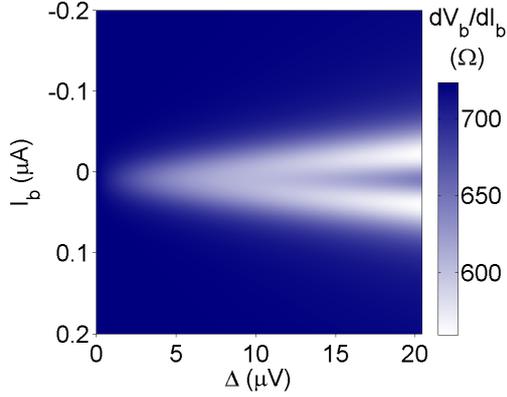

While measuring the differential contact resistance, there is no dc current applied to the Josephson junction, $I_J = 0$. So, the equivalent circuit is depicted below.

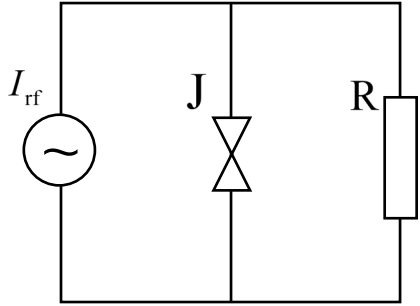

The current continuity equation becomes:

$$I_{rf0} \cos(2\pi f_{rf} t) = \frac{\hbar}{2eR} \times \frac{d\phi(t)}{dt} + I_c \sin(\phi(t)).$$

By numerically solving this differential equation, we can get the phase difference as a function of time, $\phi(t)$. The minigap of the surface state is a function of the phase difference: $\Delta(t) = \Delta_0 \left|\cos(\phi(t)/2)\right|$.[2-4]

Additionally, the rf irradiation also induces an ac bias current across the interface between the normal electrode and the surface of 3D TIs, $I'_{rf} = I'_{rf0} \cos(2\pi f_{rf} t)$. The total bias current through the interface is thus the sum of the dc bias current $I_b$ and the ac bias current $I'_{rf}$. Here, we choose $I_{rf0}/I'_{rf0} = 14.4$ to simulate the experimental data.

By now, we have obtained the minigap $\Delta(t)$ and the bias current $I_b + I'_{rf0}\sin(2\pi f_{rf} t)$ as a function of time at a given rf power and dc bias current. Then, the differential contact resistance $dV_b(t)/dI_b(t)$ at any instant time can be extracted from the R map mentioned above. The time average of $dV_b(t)/dI_b(t)$, namely $dV_b/dI_b$, can be obtained by averaging over time. By changing the dc bias current $I_b$ and the amplitude of the rf-driving ac current $I_{rf0}$, the 2D plot of the differential contact resistance as a function of bias current $I_b$ and rf power $P_{rf}$ can be obtained. Here, that we extract the differential contact resistance from the calculated R map instead of calculating it from the beginning using BTK theory, to save the calculation time.